\documentclass[12pt,a4paper,final]{iopart}
\usepackage[english]{babel}
\usepackage[utf8x]{inputenc}
\usepackage[T1]{fontenc}
\usepackage{amssymb,amsthm,amsfonts,iopams,esvect}
\usepackage{mathrsfs}
\usepackage{graphicx}
\usepackage[colorinlistoftodos]{todonotes}
\usepackage[colorlinks=true, allcolors=black]{hyperref}
\usepackage{cancel}
\usepackage{xparse}
\usepackage{braket}
\usepackage{verbatim}
\usepackage{color}
\usetikzlibrary{matrix}

\newcommand{\rrho}{\hat{\boldsymbol{\rho}}}
\newcommand{\nnu}{\hat{\boldsymbol{\nu}}}

\begin{document}

\title[Cost-effective temperature estimation with probabilistic quantum metrology]{
Cost-effective temperature estimation strategies for single-mode thermal states with probabilistic quantum metrology
}	

\author{Massimo~Frigerio$^{1,2}$, Stefano~Olivares$^{1,2}$ and Matteo~G.~A.~Paris$^{1,2}$}
\address{$^1$ Quantum Technology Lab $\&$ Applied Quantum Mechanics Group, Dipartimento di Fisica ``Aldo Pontremoli'', Universit\`a degli Studi di Milano, I-20133 Milano, Italy}
\address{$^2$ INFN, Sezione di Milano, I-20133 Milano, Italy}

\begin{abstract}
In probabilistic quantum metrology, one aims at finding weak measurements that concentrate the Fisher Information on the resulting quantum states, post-selected according to the weak outcomes. Though the Quantum Cramér-Rao bound itself cannot be overshot this way, it could be possible to improve the information-cost ratio, or even the total Fisher Information. 
We propose a post-selection protocol achieving this goal based on single-photon subtraction onto a thermal state of radiation yielding a greater information-cost ratio for the temperature parameter with respect to the standard strategy required to achieve the Quantum Cramér-Rao bound. We address just fully-classical states of radiation: this contrasts with (but does not contradict) a recent result proving that, concerning unitary quantum estimation problems, post-selection strategies can outperform direct measurement protocols only if a particular quasiprobability associated with the family of parameter-dependent quantum states becomes negative, a clear signature of nonclassicality.  
\end{abstract}
\maketitle
\section{Introduction and motivation}
In the field of local quantum estimation theory, the Quantum Cramér-Rao bound, expressed in terms of the Quantum Fisher Information (QFI), gives the ultimate bound to the precision with which one can estimate without bias a parameter that describes a given quantum state \cite{helstrom,PARIS2009,Giovannetti2011}. The standard picture involves a state $\rrho_{\lambda'}$ sampled from a one-parameter continuous family of quantum states labelled by the values of $\lambda$. The quantum estimation problem aimed at finding the value $\lambda'$ by measuring several copies of this state and constructing an unbiased estimator out of the measurement outcomes involves an optimization over all possible measurement strategies, i.e. all positive operator-valued measures (POVMs) on the Hilbert space of the system, whose solution is provided in terms of the symmetric logarithmic derivative. \\

It is then reasonable to wonder whether this bound can be exceeded by more elaborate measurement strategies that cannot be described by a (parameter-independent) POVM on the initial state, as for example weak measurements implemented by coupling the system with an ancilla and then measuring the ancilla in such a way that the conditional state of the system, post-selected according to the outcome of the weak measurement, gives an higher QFI for the parameter, as if the post-selection were able to concentrate the information in the final state: this is the framework of \emph{probabilistic quantum metrology} \cite{probmetr06,probmetr13,probmetr16,probmetr162,postselectcost,Chen2021}. However, general results show that these strategies cannot overcome the full Quantum Cramér-Rao bound \cite{weakmeas13,weakmeas2015,weakmeassubopt,cramerraopostselect}, ultimately due to a compensation between the enhanced QFI of the post-selected state and the probability of success of the post-selection process; moreover, the same conclusion holds true if one considers all the information available from the strategy, i.e. by adding to the count the Fisher Information of the weak measurement's outcomes and the QFI of the discarded states of the system \cite{weakmeas14}. On the other hand, the optimal measurement specified by the symmetric logarithmic derivative, despite fixing the ultimate quantum bound, is often unreasonable to implement in an experimental setting, which is the natural purpose of quantum metrology. Thus, at least two loopholes open in the above argument. Firstly, probabilistic quantum metrology may actually achieve an advantage over the direct measurement procedure when sub-optimal, realistic measurement strategies are employed; this has been studied, e.g., in the context of the estimation of the gain of a nondeterministic linear amplifier \cite{Adnane2019} and in quantum magnetometry with nitrogen-vacancy centers of diamond \cite{Coto2021}. Secondly, taking into account the cost of different detection methods, the overall information-cost ratio could be drastically improved by deciding to spend a higher amount of resources only in the trials with a successful post-selection outcome. \\

\indent In this paper, we provide a detailed example to show that such advantages can indeed happen even when only fully classical states are involved, thereby excluding the necessity of \emph{quantumness} to render post-selection profitable. We consider a common de-Gaussianification protocol, the single-photon subtraction, both with an ideal model and with a realistic implementation, applied to a thermal state of a single mode of radiation. We show that the ideal photon subtraction yields a non-Gaussian state with a higher QFI with respect to the temperature parameter, suggesting a chance to overcome the Quantum Cramér-Rao bound. However, single-photon subtraction is not a unitary operation. A standard experimental realization of the single-photon subtraction, which is easily implemented with ordinary quantum optics equipment, involves sending the quantum state onto a beam splitter with very high transmittance and performing an on-off measurement on the weak, reflected beam: whenever the on-off detector clicks, it is very likely that a single photon has been detected and the transmitted state will approximate the photon-subtracted state. Thus, this is precisely an instance of probabilistic quantum metrology. Despite the no-go results discussed above, we show that already in this basic framework the two alternatives allow for an advantage over the standard strategy: namely, if an heterodyne measurement protocol is assumed in place of the optimal photon-counting measurement needed to reach the QFI, the conditional strategy improves the total Fisher Information for the temperature parameter. Moreover, taking into account the measurement costs, the FI-to-resources ratio can dramatically improve with the realistic single-photon subtraction protocol.

\section{Photon subtracted thermal states}
We begin by considering a thermal state of a single mode \cite{OLIVARES2021127720}, characterized by an average number of photons $\lambda = ( e^{ \beta} - 1 )^{-1}$, where $\beta =  \hbar \omega / (k_{B} T)$, $T$ is the temperature parameter of the state and $\omega$ is the frequency of the mode. In the number basis, the state can be expressed as:
\begin{eqnarray}
    \label{eq:defTH}
     \nnu_{\lambda}  &:= \ \sum_{n=0}^{+\infty}  p_{n}(\lambda) \ \vert n \rangle \langle n \vert  \,, \\
     p_{n}(\lambda)  \ &:= \ \frac{1}{1+ \lambda} \left( \frac{\lambda}{1+\lambda} \right)^{n} \,.
\end{eqnarray}
The value of $\lambda$ will also play the role of the parameter to be estimated. It is worth mentioning that this state would describe the photon statistics of a single mode of radiation in thermal equilibrium \emph{if it was emitted from a perfectly coherent source}. 

\subsection{Ideal one-photon subtraction}
Ideally, the subtraction of a single photon is implemented on the Hilbert space by the action of the lowering operator $\hat{a}$ of the corresponding mode. If we denote ${\nnu^{-}}_{\lambda}$ the ideal one-photon subtracted thermal state, i.e.:
\begin{equation}
    {\nnu^{-}}_{\lambda}  :=  \mathcal{E}^{-} \left[ \nnu_{\lambda} \right]  =  \frac{\hat{a}  \nnu_{\lambda} \hat{a}^{\dagger} }{ \Tr[ \nnu_{\lambda} \hat{a}^{\dagger} \hat{a}  ] }  =  \frac{1}{\lambda}\hat{a}  \nnu_{\lambda} \hat{a}^{\dagger}  \,.
\end{equation}
its photon-number probability distribution will be:
\begin{equation}
\label{eq:PnPhSubt}
p^{-}_{n}(\lambda) \ \ = \ \ \frac{n+1}{(1+\lambda)^{2}} \left( \frac{\lambda}{1+\lambda} \right)^{n}   \,.
\end{equation}
Notice that the average number of photons in the ideal one-photon subtracted thermal state is:
\[ \Tr [ \nnu^{-}_{\lambda} \hat{a}^{\dagger} \hat{a} ] \ = \  2 \lambda \]
and, perhaps counter-intuitively, this is greater than $\lambda$, the average number of photons in the starting thermal state, before the ideal one-photon subtraction. This is peculiar of the thermal distribution, and it can be partially understood by acknowledging that the vacuum state is always the most probable number state in a thermal distribution, and it is removed by the one-photon subtraction map. Notice also that the one-photon subtraction map does not preserve the energy. \\

We shall need to consider also other probability distributions associated with ${\nnu}_\lambda$ and $\nnu^{-}_{\lambda}$, whose expressions are easier to derive starting from an operator representation of these states. It is well known that a single mode thermal state can also be written as:  
\begin{equation}
\label{eq:THOp}
    \nnu_{\lambda} \ = \ (1 - e^{-\beta}) e^{-\beta \hat{a}^\dagger \hat{a}}  \,.
\end{equation}
Applying the ideal one-photon subtraction map $\mathcal{E}^{-}$ to this expression, we get:
\begin{equation}
\label{eq:subtOp}
    \nnu^{-}_{\lambda} \ = \ (1- e^{-\beta})^{2} \left[ \mathbb{I} + \hat{a}^\dagger \hat{a} \right]  e^{-\beta \hat{a}^\dagger \hat{a}} \ = \  (1- e^{-\beta})^{2} \left[ \mathbb{I} - \frac{ \partial}{\partial \beta} \right] e^{-\beta \hat{a}^\dagger \hat{a}}   \,.
\end{equation}
The last expression is very convenient to readily evaluate the homodyne and heterodyne distributions of $\nnu^{-}_{\lambda}$ from the well-known corresponding distributions of the thermal state $\nnu_{\lambda}$.\\

In a completely analogous way, the one-photon added thermal state defined by:
\begin{equation}
\label{eq:addOp}
    {\nnu^{+}}_{\lambda} \ :=  \mathcal{E}^{+} \left[ \nnu_{\lambda} \right] \ =  \  \frac{1}{\lambda + 1}\hat{a}^{\dagger}  \nnu_{\lambda} \hat{a}  \,,
\end{equation}
can be rewritten as:
\begin{equation}
    \nnu^{+}_{\lambda} \ =  \ (e^{\beta} - 1)(1- e^{-\beta}) \  \hat{a}^\dagger \hat{a} \   e^{-\beta \hat{a}^\dagger \hat{a}} \ = \   -(e^{\beta} - 1)(1- e^{-\beta}) \frac{ \partial}{\partial \beta}  e^{-\beta \hat{a}^\dagger \hat{a}}   \,.
\end{equation}
The corresponding photon-number distribution is:
\begin{equation}
\label{eq:PnPhAdd}
p^{+}_{n}(\lambda) \ \ = \ \ \frac{n}{\lambda (1+\lambda)} \left( \frac{\lambda}{1+\lambda} \right)^{n}  \,.
\end{equation}
and the average photon number is $2\lambda + 1$ in this case. \\

The Husimi Q-function of these states are easily evaluated from Eq.(\ref{eq:THOp}), Eq.(\ref{eq:subtOp}) and Eq.(\ref{eq:addOp}) with the help of the definition:
\begin{equation}
    Q \left[ \rrho \right] ( \alpha ) \ \ := \ \ \frac{1}{\pi} \langle \alpha \vert \rrho \vert \alpha \rangle \ , \ \ \ \ \ \ \ \alpha \in \mathbb{C} \ \,.
\end{equation}
The results are as follows:
\begin{eqnarray}
    Q_{\lambda} (\alpha) \ \ &= \ \ \frac{1}{\pi ( 1+ \lambda)} \exp \left[- \frac{ \vert \alpha \vert^{2} }{1+ \lambda} \right] \,, \\
    Q^{-}_{\lambda} (\alpha) \ \ &= \ \ \frac{ 1 + \lambda(1 +  \vert \alpha \vert^{2} }{ \pi ( 1+ \lambda)^{3}}   \exp \left[ - \frac{  \vert \alpha \vert^{2} }{1+ \lambda} \right]  \,, \\
    Q^{+}_{\lambda} (\alpha) \ \ &= \ \ \frac{   \vert \alpha \vert^{2}  }{ \pi ( 1+ \lambda)^{2}}   \exp \left[ - \frac{ \vert \alpha \vert^{2} }{1+ \lambda} \right]   \,.
\end{eqnarray}

In a completely analogous manner, it is possible to derive the probability distributions associated with the quadrature $\hat{x} = \frac{ \hat{a} + \hat{a}^{\dagger}}{\sqrt{2}}$ on the states above. The results are:
\begin{eqnarray}
    X_{\lambda} (x) \  &= \ \frac{1}{\sqrt{\pi} \sqrt{ 1+ 2\lambda}} \exp \left[- \frac{ x^{2} }{1+ 2\lambda} \right]  \,, \\
    X^{-}_{\lambda} (x) \  &= \ \frac{ 1 + \lambda( 3 + 2x^2 + 2 \lambda)}{\sqrt{\pi} ( 1+ 2\lambda)^{5/2}}   \exp \left[ - \frac{ x^{2} }{1+ 2\lambda} \right] \,, \\
    X^{+}_{\lambda} (x) \  &= \ \frac{ \lambda + 2 (\lambda^{2} + x^{2}(1+\lambda) ) }{\sqrt{\pi} ( 1+ 2\lambda)^{5/2}}   \exp \left[ - \frac{ x^{2} }{1+ 2 \lambda} \right]  \,.
\end{eqnarray}
It is worth mentioning that, among Gaussian measurements, either heterodyne detection (whose statistics is described by the Husimi Q-function) or homodyne detection are provably optimal to estimate the temperature of thermal states, as recently discussed in \cite{cenni2021thermometry}. However, it is not obvious if this holds for the photon-subtracted and photon-added thermal states.

\subsection{Realistic one-photon subtraction}
Photon-subtraction is a cornerstone of many quantum optical protocols, and it is routinely exploited to prepare nonclassical and non-Gaussian states of light \cite{Our06,Our07,Weng04,Fiu05,Oli05}, as well as to test fundamental aspects of quantum mechanics \cite{Par07,Par08,Oli04,Inv05} and to improve quantum teleportation with continuous variables \cite{Opa00,Oli03}. The standard way to implement approximate one-photon subtraction on a state in quantum optics involves sending it onto a high transmittance beam-splitter, with the vacuum state in the orthogonal port, and performing on-off detection on the weak, reflected signal \cite{Barnett18}. In the limit of infinitesimal reflectivity of the beam-splitter and perfect efficiency of the on-off detection, whenever the detectors catches a signal, it will be a single photon and the transmitted, output state will be one-photon subtracted according to the ideal map $\mathcal{E}^{-}$ introduced above. \\

In a realistic scenario, the action of the beam-splitter will be described by the unitary operator:
\begin{equation}
\label{eq:UBSstand}
    \hat{U}_{\mathrm{BS}} (\zeta ) \ \ \ := \ \ \ \exp \left[ \zeta \hat{a}^\dagger \hat{b} - \zeta^{*} \hat{a} \hat{b}^{\dagger} \right]  \,.
\end{equation}
where $\hat{a}$ is the mode operator of the field we are interested in and $\hat{b}$ is the mode operator of the orthogonal port. The parameter $\zeta = \theta e^{i \phi}$ quantifies the transmittance and the additional phase introduced by the beam-splitter. 
To explicitly carry out the calculations for generic values of beam-splitter transmittance and detection efficiency, it is convenient to express the beam-splitter unitary evolution in the following form:
\begin{equation}
\label{eq:UBS}
      \hat{U}_{\mathrm{BS}} (\zeta ) \  =  \ \exp \left[ -e^{-i\phi} \tan \theta \ \hat{a} \hat{b}^{\dagger} \right]  \left( \cos^{2} \theta \right)^{\frac{1}{2} (\hat{a}^\dagger \hat{a} - \hat{b}^\dagger \hat{b} )} \exp \left[ e^{i\phi} \tan \theta \  \hat{a}^{\dagger} \hat{b} \right]  \,.
\end{equation}
which can be derived from the standard expression Eq.(\ref{eq:UBSstand}) by exploiting the rules of the two-boson representation of the $\mathfrak{su}(2)$ Lie algebra provided by the operators $\hat{J}_{+} = \hat{a}^{\dagger} \hat{b}$, $\hat{J}_{-} = \hat{a} \hat{b}^{\dagger}$ and $\hat{J}_{3} = \frac{1}{2} [ \hat{J}_{+}, \hat{J}_{-} ] = \frac{1}{2} ( \hat{a}^\dagger \hat{a} - \hat{b}^\dagger \hat{b} ) $. For our realistic one-photon subtraction protocol we will assume $\phi = 0$ and $\eta = \cos^{2} \theta$ will be the beam-splitter \emph{transmittance}. If $\theta = 0$, then $\eta = 1$ and $\hat{U}_{\mathrm{BS}} = \mathbb{I}$, so that the input state is fully transmitted.\\

As an input state, we shall take $\nnu_{\lambda} \otimes \vert 0 \rangle \langle 0 \vert$, which is simply the thermal state to be analyzed on the first mode and the vacuum state in the other mode. It is now a simple matter to write down the output state of the two-mode after the beam-splitter:

\begin{eqnarray}
\label{eq:RBS1}
     &\hat{U}_{\mathrm{BS}}(\theta) \left[ \nnu_{\lambda} \otimes \vert 0 \rangle \langle 0 \vert \right] \hat{U}^{\dagger}_{\mathrm{BS}}(\theta) \  
    = \ \frac{1}{1+\lambda} \sum_{n=0}^{+\infty} \left( \frac{ \lambda}{1+\lambda} \right)^{n} \times \\
    & \times \sum_{k,k'=0}^{n} \sqrt{ {n\choose k} {n\choose k'} } (1-\eta)^{\frac{k+k'}{2}} \eta^{n- \frac{k+k'}{2}}  
     \vert n - k \rangle \langle n- k' \vert \otimes \vert k \rangle \langle k' \vert \nonumber   \,.
\end{eqnarray}

On this state, an on-off detection with efficiency $0 < \epsilon \leq 1$ is performed on the second mode. This is implemented at the mathematical level by a POVM with the following two operators:
\begin{equation}
\label{eq:YNPOVM1}
    \hat{\Pi}_{0} \ = \ \sum_{m=0}^{+\infty} ( 1 - \epsilon )^{m} \vert m \rangle \langle m \vert  \,,
\end{equation}
\begin{equation}
\label{eq:YNPOVM2}
    \hat{\Pi}_{1} \ = \ \mathbb{I} - \hat{\Pi}_{0} \ = \ \sum_{m=0}^{+\infty} \left[ 1 - (1 -\epsilon)^{m} \right] \vert m \rangle \langle m \vert  \,.
\end{equation}
Whenever the detector clicks, meaning that the outcome is $1$, the state of the first mode will be the approximate one-photon subtracted thermal state. This happens with probability:
\begin{eqnarray}
\label{eq:paccsubt}
    \wp_{1}(\lambda ; \eta, \epsilon ) \ & = \ \Tr \left[ \left( \mathbb{I} \otimes \hat{\Pi}_{1} \right)   \hat{U}_{\mathrm{BS}}(\theta) \big[ \nnu_{\lambda} \otimes \vert 0 \rangle \langle 0 \vert \big] \hat{U}^{\dagger}_{\mathrm{BS}}(\theta) \right] \ = \\ 
     &= \  \frac{ \lambda \epsilon (1- \eta)}{1 +\lambda \epsilon (1- \eta) }   \,. \nonumber
\end{eqnarray}
where we used a different calligraphic style for the letter "p" to distinguish it from the photon-number distribution of the initial thermal state. If we denote by $\rrho_{\lambda}$ the approximate one-photon subtracted thermal state and by $\Tr_{B}$ the partial trace over the second output mode of the beam-splitter, we can finally write:
\begin{eqnarray}
\label{eq:PnBSPhSubt}
    \rrho_{\lambda} \ &=  \ \frac{1}{\wp_{1}} \Tr_{B} \left[ \left( \mathbb{I} \otimes \hat{\Pi}_{1} \right)   \hat{U}_{\mathrm{BS}}(\theta) \big[ \nnu_{\lambda}  \otimes \vert 0 \rangle \langle 0 \vert \big] \hat{U}^{\dagger}_{\mathrm{BS}}(\theta) \right]  \ =  \\
    & = \frac{1}{\wp_{1}} \frac{1}{1 + \eta \lambda} \sum_{n=0}^{+\infty} \left( \frac{\eta \lambda}{1+ \eta \lambda} \right)^{n} \left[ 1 - \left( \frac{ 1 + \eta \lambda}{1 + \lambda(\eta + (1-\eta)\epsilon)} \right)^{n+1} \right] \vert n \rangle \langle n \vert \nonumber \,.
\end{eqnarray}

The steps to arrive at this expression are easily reconstructed by rearranging the sums and applying the binomial theorem. Eq.(\ref{eq:PnBSPhSubt}) has several interesting features: apart from the factor in square brackets and the normalization, the usual distribution of a thermal state with average number of photons $\eta \lambda$ can be clearly recognized: this would be the state of the first mode after the beam-splitter if no detection took place on the reflected beam. Moreover, both Eq.(\ref{eq:RBS1}) and the final result in Eq.(\ref{eq:PnBSPhSubt}) involve P-classical states, in the sense of Glauber \cite{glauber63,sudarsh63,cahillglauber69,glauber69}. Indeed, the former is the result of a P-classical state mixed with the vacuum by a beam-splitter, while the final state is diagonal in the number basis and with non vanishing photon-number probability for any $n \geq 0$, hence it must be P-classical too; this also implies that their Wigner functions are everywhere non-negative in phase space. However, the state in Eq.(\ref{eq:PnBSPhSubt}) is non-Gaussian, having a Wigner function which is a linear combination of two Gaussians in phase space. This is due to the fact that, although the operator $\hat{\Pi}_{0}$ associated with the rejection of the selected state is Gaussian, the operator $\hat{\Pi}_{1}$ corresponding to the accepted state is not.

\section{Fisher Information for one-photon subtracted thermal states}
As a preliminary remark, we highlight the fact that all the states considered above are diagonal in the number basis, as in Eqs.(\ref{eq:defTH}), (\ref{eq:PnPhSubt}), (\ref{eq:PnPhAdd}), and (\ref{eq:PnBSPhSubt}), whose vectors are independent of $\lambda$; therefore, the QFI of these states for the parameter $\lambda$ reduces to the (classical) FI of their photon-number probability distributions. In other words, the problem of estimating $\lambda$ on these states is a classical estimation problem embedded in a quantum setting. In this sense, single-photon detection is a \emph{classical} detection strategy for this special scenario. Also, no entanglement is created by the beam-splitter, precisely because the input state is P-classical. \\

\begin{figure}
\centering
\includegraphics[width=0.7 \textwidth]{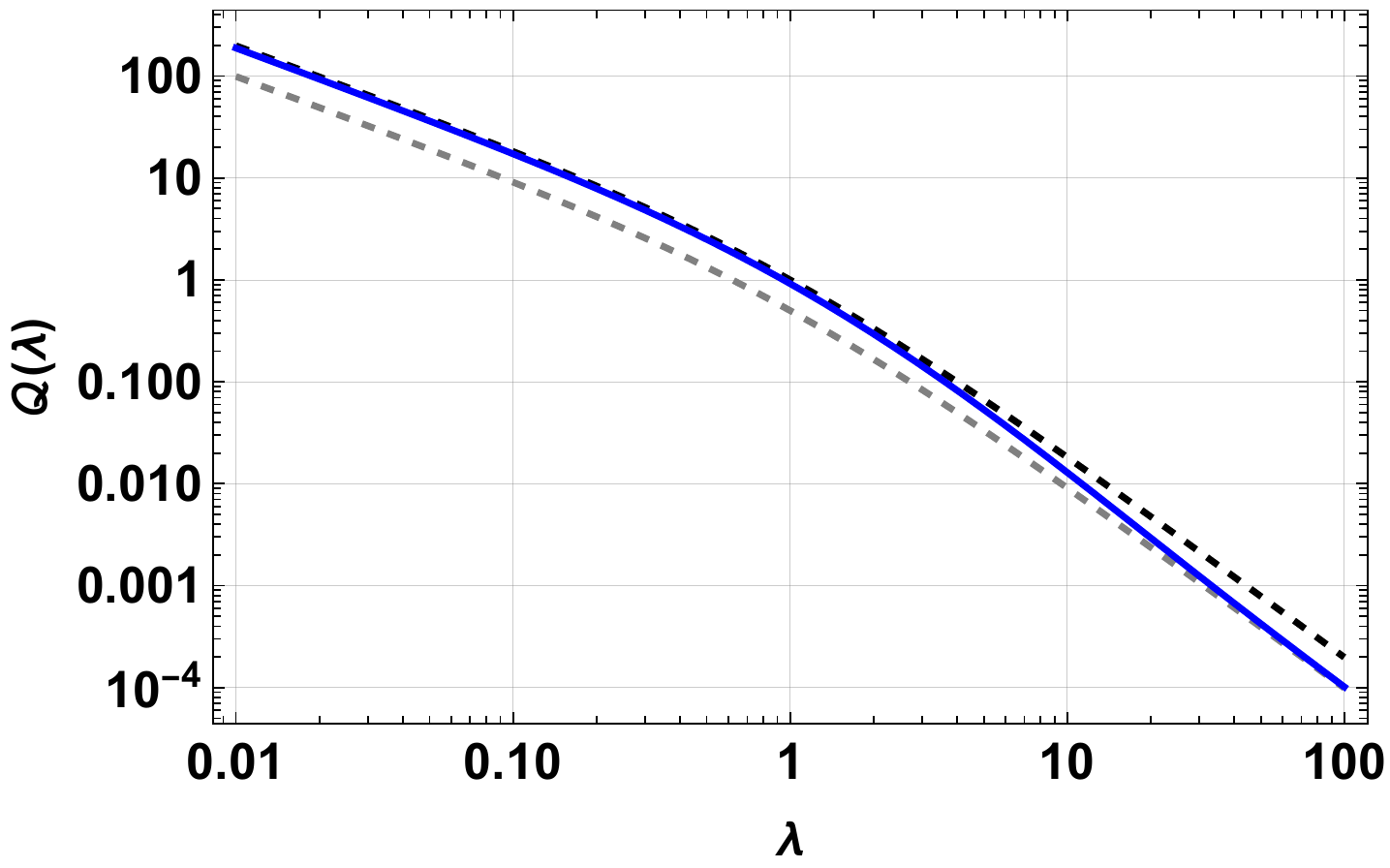}
$\quad $
\caption{\label{fig:QFIThComp}
Quantum Fisher Information vs. initial average photon number $\lambda = (e^{\beta} - 1)^{-1}$ for thermal state (dashed gray), ideal one-photon subtracted / added thermal state (solid black) and realistic one-photon subtracted thermal state (blue) for a beam-splitter with transmittance $\eta = 0.95$ and on-off detection with efficiency $\epsilon = 0.99$. Mind the log-scale on both axes.}
\end{figure}

With the help of the standard formula for the FI for a parameter $\lambda$ encoded in a probability distribution $P_{k}(\lambda)$ over a discrete random variable $k$:
\begin{equation}
    \mathcal{F}[ \{ P_{k}(\lambda) \} ] \ \ := \ \ \sum_{k} P_{k}(\lambda) \left[ \partial_{\lambda} \log P_{k} (\lambda) \right]^{2}  \ \ = \ \  \sum_{k}  \frac{ \left( \partial_{\lambda} P_{k} (\lambda) \right)^{2}}{ P_{k}(\lambda)}  \,.
\end{equation}
we arrived at the following expressions for the Quantum Fisher Information $\mathcal{Q}$ of $\nnu_{\lambda}, \nnu^{-}_{\lambda}$ and $\nnu^{+}_{\lambda}$:
\begin{equation}
    \label{eq:QFITh} \mathcal{Q} \left[ \nnu_{\lambda} \right] \ = \ \mathcal{F} \left[ p_{n} (\lambda) \right] \ = \ \frac{1}{\lambda( 1 + \lambda)}  \,, 
  \end{equation}
\begin{equation} 
\label{eq:QFISubt} \mathcal{Q} \left[ \nnu^{-}_{\lambda} \right] \ = \  \mathcal{F} \left[ p^{-}_{n} (\lambda) \right] \ = \ \frac{2}{\lambda( 1 + \lambda)} \ = \ \mathcal{Q} \left[ \nnu^{+}_{\lambda} \right]   \,.
\end{equation}
Thus we see that both ideal one-photon subtraction and addition double the QFI. The same computation for the more involved photon-number distribution of the realistic one-photon subtracted thermal state, Eq.(\ref{eq:PnBSPhSubt}) leads to a series that we were not able to sum in a closed form. However, by performing the sum numerically and plotting the resulting QFI as a function of $\lambda$ against the results of Eq.(\ref{eq:QFITh},\ref{eq:QFISubt}), we confirmed that the advantage in the QFI can still be attained by the realistic one-photon subtraction protocol, as shown in Fig.(\ref{fig:QFIThComp}) for plausible values of beam-splitter transmittance and detector efficiency ($\eta = 0.95$ and $\epsilon = 0.99$).  

It can be seen that the QFI of the realistic one-photon subtraction converges to the ideal case in the limit $\lambda \to 0$: indeed, when the initial thermal state has a small average number of photons, the realistic protocol well-approximates the ideal scenario, because whenever the detector clicks, there is a high chance that just a single photon was actually caught. However, this is in spite of the probability of success $\wp_{1}(\lambda)$ of the protocol itself, and we shall consider this issue in a later section. When $\lambda \gg 1$, instead, the realistic one-photon subtracted thermal state performs no better than the initial thermal state itself.

\section{Exploring other cost-effective measurement strategies}
Despite the fact that the number basis is the one that translates our estimation task on thermal quantum states into a classical one, actually resolving the photon-number distribution is a formidable experimental work. Hence, it is interesting to study whether the advantage over the QFI of the unprocessed thermal state persist if we consider easier-to-implement measurements on the ideal and realistic one-photon subtracted thermal states. \\

We considered three detection strategies:
\begin{enumerate}
    \item \textbf{\emph{Homodyne detection}}  of the $\hat{x}$ quadrature, a standard detection protocol in quantum optics that involves combining the state with a reference coherent state (local oscillator) onto a 50/50 beam splitter and evaluating the difference bewteen the photocurrents at the two outputs
    \item \textbf{\emph{Heterodyne detection}} (or \emph{double homodyne detection}) which effectively implements a POVM whose elements are coherent-state projectors
    \item \textbf{\emph{On-off detection}} with efficiency $\epsilon$, described by the POVM of Eq.(\ref{eq:YNPOVM1},\ref{eq:YNPOVM2}) as before
\end{enumerate}

\begin{figure}[h!]
\centering
\includegraphics[width=0.7 \textwidth]{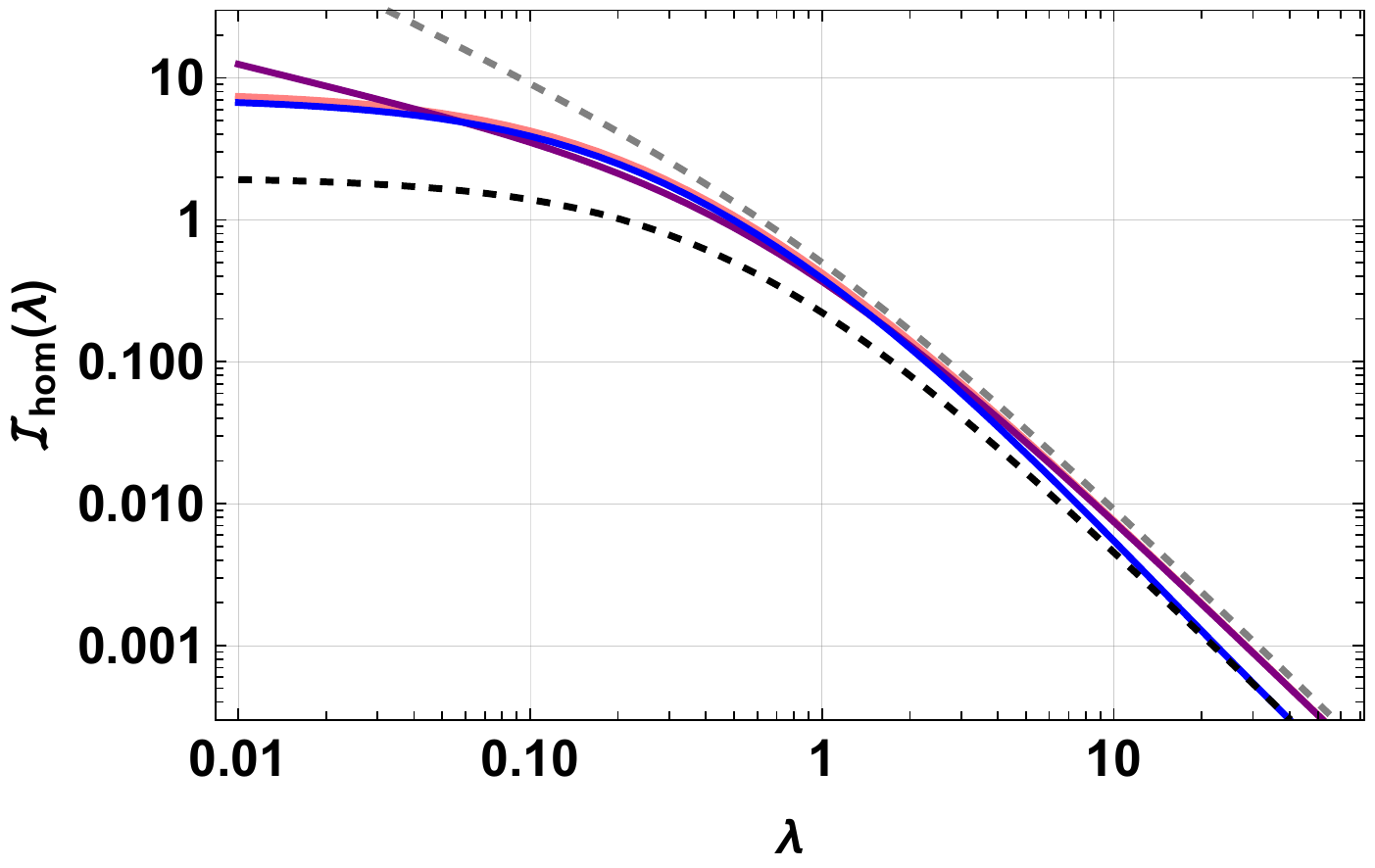}
$\quad $
\caption{\label{fig:FIHomComp}
 Fisher Information vs. initial average photon number $\lambda = (e^{\beta} - 1)^{-1}$ of homodyne distributions for thermal state (dashed black), ideal one-photon subtracted / added thermal state (solid pink / purple) and realistic one-photon subtracted thermal state (solid blue) for a beam-splitter with transmittance $\eta = 0.95$ and on-off detection with efficiency $\epsilon = 0.97$. Mind the log-scale on both axes. Quantum Fisher Information of the initial thermal state (dashed gray) for reference.}
\end{figure}

Starting with homodyne measurements, Fig.(\ref{fig:FIHomComp}) shows that it always performs worse than the QFI of the initial thermal state. However, the FI of the homodyne distributions $X^{-}_{\lambda}$, $X^{+}_{\lambda}$ and $\langle x \vert \rrho_{\lambda} \vert x \rangle$ are greater than the FI of $X_{\lambda}$ over a wide range of values of $\lambda$, while for $\lambda \gg 50$ the approximate one-photon subtracted thermal state provides less information than the initial thermal state under homodyne detection. 

\begin{figure}[h!]
\centering
\includegraphics[width=0.7 \textwidth]{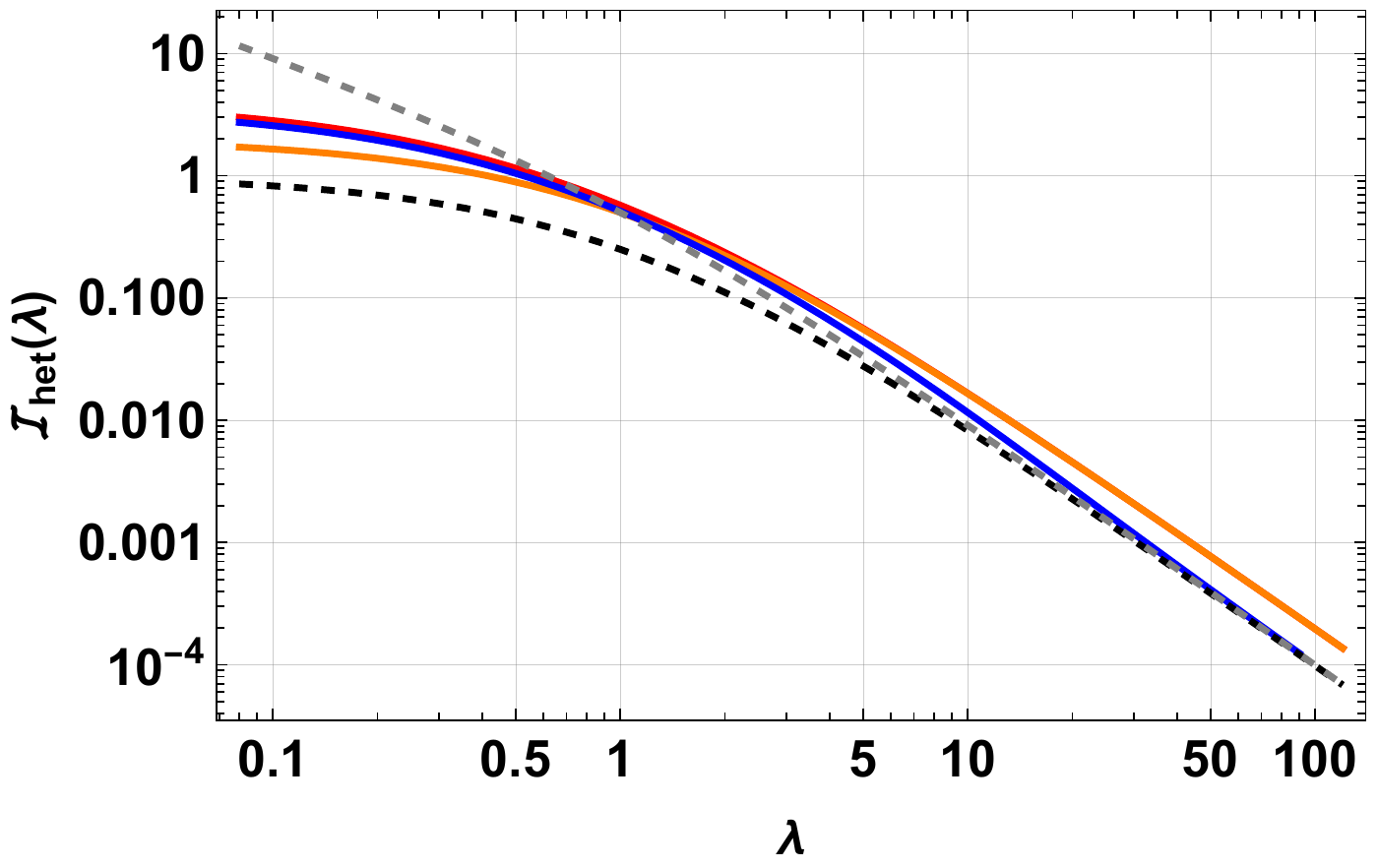}
$\quad $
\caption{\label{fig:FIHetComp}
 Fisher Information vs. initial average photon number $\lambda = (e^{\beta} - 1)^{-1}$ of heterodyne distributions for thermal state (dashed black), ideal one-photon subtracted / added thermal state (solid red / orange) and realistic one-photon subtracted thermal state (solid blue) for a beam-splitter with transmittance $\eta = 0.95$ and on-off detection with efficiency $\epsilon = 0.97$. Mind the log-scale on both axes. Quantum Fisher Information of the initial thermal state (dashed gray) for reference.}
\end{figure}

Considering heterodyne detection instead, we found that $\nnu^{\pm}_{\lambda}$ and $\rrho_{\lambda}$ outperform even the QFI of the initial thermal state, in the range $\lambda \gtrsim 1$, while they perform better than the heterodyne distribution of $\nnu_{\lambda}$ for a greater range of values, but $\rrho_{\lambda}$ eventually becomes worse than $\nnu_{\lambda}$ for this detection strategy. These trends are displayed in Fig.(\ref{fig:FIHetComp}).

\begin{figure}
\centering
\includegraphics[width=0.7 \textwidth]{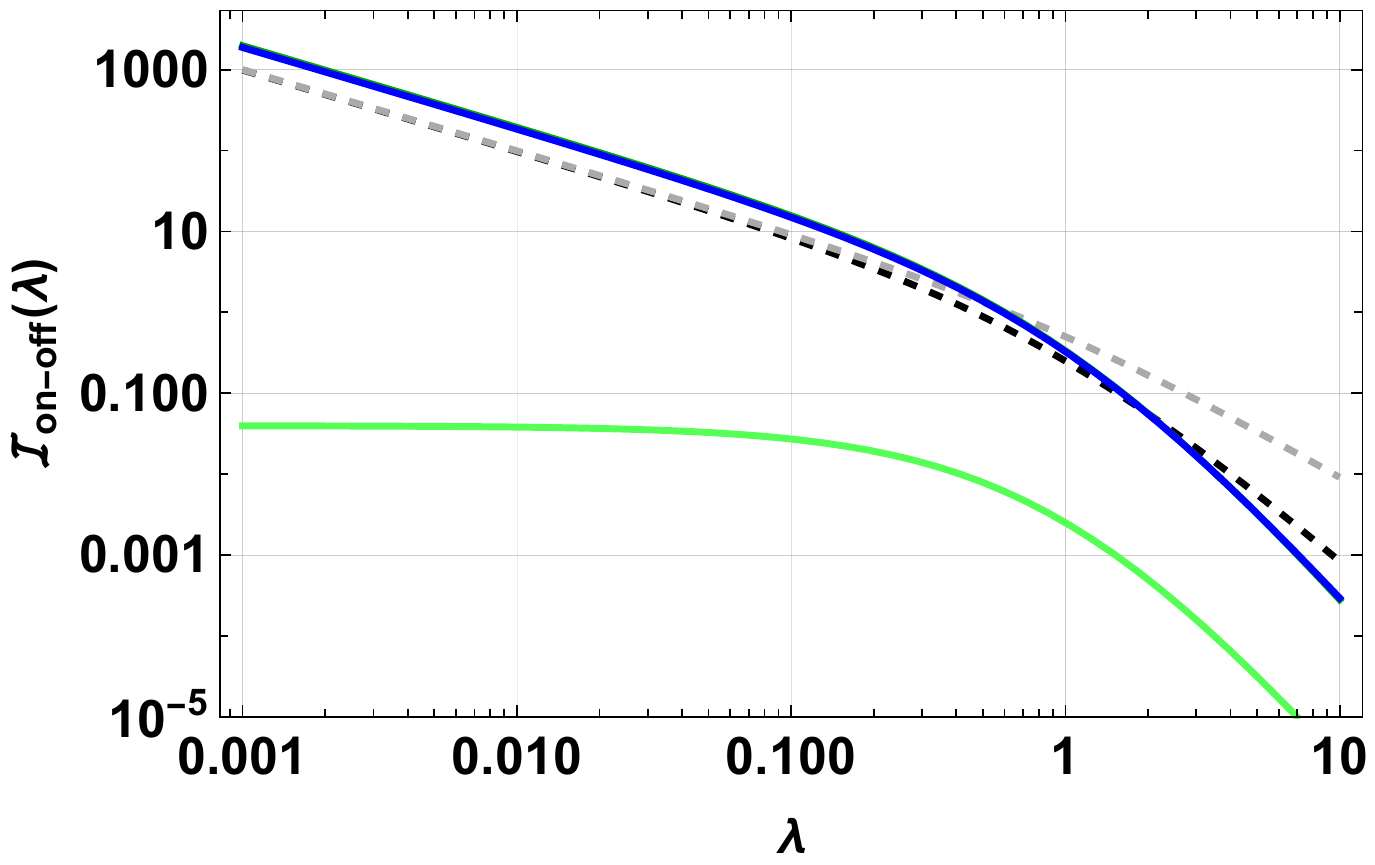}
$\quad $
\caption{\label{fig:FIYNComp}
 Fisher Information vs. initial average photon number $\lambda = (e^{\beta} - 1)^{-1}$ of on-off distributions for thermal state (dashed black), ideal one-photon subtracted (solid blue curve), photon-added thermal state (solid light-green) and realistic one-photon subtracted thermal state (solid blue curve, coinciding with the ideal photon-added state in this range) for a beam-splitter with transmittance $\eta = 0.95$ and on-off detection with efficiency $\epsilon = 0.97$. Mind the log-scale on both axes. Quantum Fisher Information of the initial thermal state (dashed gray) for reference.}
\end{figure}

Finally, let us discuss the cheapest strategy, on-off detection. With the idea of exploiting a resource we already harnessed, we will assume the detection efficiency to be the same of the on-off detection employed in the realistic one-photon subtraction protocol. In that case, the realistic scenario can be summarized by a black box with one input for the initial thermal state and two output for the signal of the two identical on-off detectors, put at the two ends of the beam-splitter. The signal of the first detector is recorded whenever the second detector clicks, meaning that one-photon subtraction happened. This can result in a relatively compact and easy-to-build apparatus. Fig.(\ref{fig:FIYNComp}) shows that the FI of the on-off distributions for $\nnu^{-}_{\lambda}$ and $\rrho_{\lambda}$ are significantly greater than the QFI of the initial thermal state for $\lambda \lesssim 0.5$ and are still better than the FI of the on-off distribution of $\nnu_{\lambda}$ for $\lambda \lesssim 2$. On the other hand, the ideal photon-added state $\nnu^{+}_{\lambda}$ is completely inefficient with this detection strategy; the reason can be seen in the fact that the ideal photon-added state, unlike the initial thermal state, will almost always make an efficient on-off detector click, because it has no contribution from the vacuum state; therefore the on-off statistics will depend very weakly on the initial parameter of the thermal distribution.

\section{Exploiting all the information from the realistic single-photon subtraction protocol}
\begin{figure}
\centering
\includegraphics[width=0.7 \textwidth]{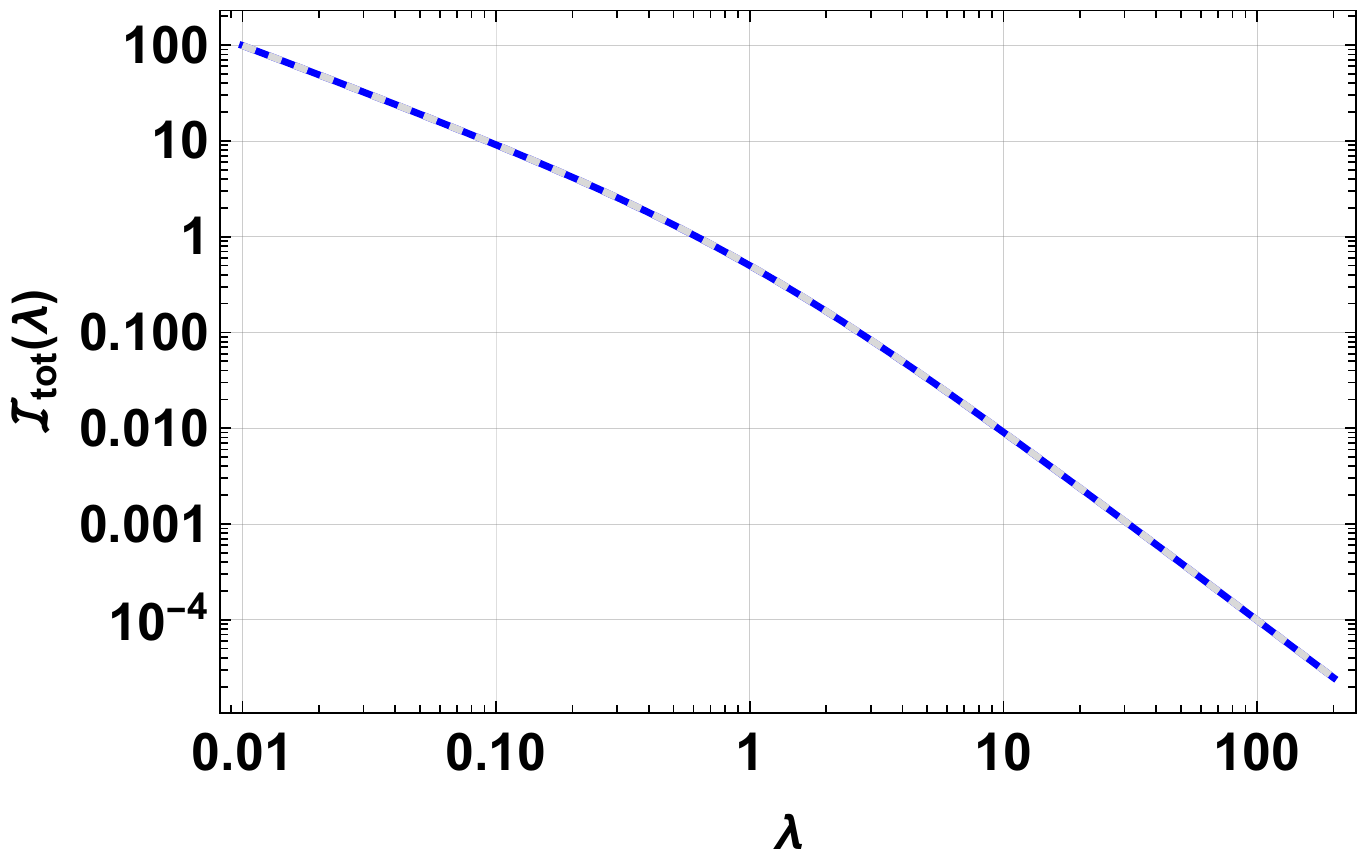}
$\quad $
\caption{\label{fig:QFItotal}
 Total information $\mathcal{F}_{\mathrm{tot}}$ on $\lambda$ extractable from the realistic one-photon subtraction protocol on $\nnu_{\lambda}$ (solid blue) for a beam-splitter with transmittance $\eta = 0.95$ and on-off detection with efficiency $\epsilon = 0.97$, vs. QFI of the original thermal state $\nnu_{\lambda}$ (dashed gray). }
\end{figure}

In evaluating the Fisher Information of the probability distributions associated with the realistic one-photon subtracted thermal state, we disregarded a crucial issue: the realistic protocol is not always successful. If we repeat the protocol $M$ times on identically prepared thermal states $\nnu_{\lambda}$, according to Eq.(\ref{eq:paccsubt}) we will prepare:
\begin{equation}
    M' \ \ = \ \ M \wp_{1} (\lambda ) \ \ = \ \ \frac{ M \lambda \epsilon (1 - \eta)}{1 + \lambda \epsilon (1-\eta)}  \,.
\end{equation}
approximate one-photon subtracted thermal states $\rrho_{\lambda}$. Since the Cramér-Rao relation writes:
\begin{equation}
    \mathrm{Var} \big[ \hat{\lambda} \big] \ \ \geq \ \ \frac{1}{M' \ \mathcal{F} [ \rrho_{\lambda} ]} \ \ = \ \ \frac{1}{M \wp_{1} (\lambda) \mathcal{F} [ \rrho_{\lambda} ]}  \,.
\end{equation}
for the variance of any unbiased estimator $\hat{\lambda}$ for the parameter $\lambda$, we deduce that having a reduced number of accepted one-photon subtracted thermal states effectively translates into a reduction of the Fisher Information extractable by any measurement, by a $\lambda$-dependent factor which is precisely the probability of success $\wp_{1}(\lambda)$. However, stopping here would yield to a pessimistic underestimate: when the subtraction protocol fails, we still have a final state that can be measured. This is given by:

\begin{eqnarray}
\label{eq:rejphotsubt}
    \rrho^{\oslash}_{\lambda} &=  \frac{1}{\wp_{0} (\lambda)}  \frac{1}{1 + \lambda(\eta + (1-\eta)\epsilon)} \sum_{n=0}^{+\infty} \left( \frac{\eta \lambda}{1 + \eta \lambda +  \lambda (1-\eta)\epsilon} \right)^{n} \vert n \rangle \langle n \vert \ = \ \ \\
    &=  \left( 1 - \frac{ \eta \lambda }{1 + \eta \lambda +  \lambda (1-\eta)\epsilon} \right) \sum_{n=0}^{+\infty} \left( \frac{\eta \lambda}{1 + \eta \lambda +  \lambda (1-\eta)\epsilon} \right)^{n} \vert n \rangle \langle n \vert \nonumber  \,.
\end{eqnarray}

where $\wp_{0} (\lambda) = 1 - \wp_{1}(\lambda) = \frac{1}{ 1 + \lambda ( 1 - \eta) \epsilon}$. By writing:
\[    \frac{ \tilde{\lambda} }{ \tilde{\lambda} + 1 } \ = \ \frac{ \eta \lambda }{1 + \eta \lambda +  \lambda (1-\eta)\epsilon}    \,,   \]
and solving for $\tilde{\lambda}$:
\begin{equation}
    \tilde{\lambda} \ = \ \frac{\eta \lambda}{ 1 + (1-\eta) \epsilon \lambda} \,,
\end{equation}
we can rewrite Eq.(\ref{eq:rejphotsubt}) as:
\begin{equation}
    \rrho^{\oslash}_{\lambda} \ \ = \ \ \frac{1}{ 1 + \tilde{\lambda}} \sum_{n=0}^{+\infty} \left( \frac{\tilde{\lambda}}{1+ \tilde{\lambda}} \right)^{n}  \vert n \rangle \langle n \vert \,.
\end{equation}
Thus, whenever the realistic one-photon subtraction protocol fails, the resulting state is again a thermal state, but with a new average photon number $\tilde{\lambda}$. The QFI with respect to $\lambda$ of $\nnu_{\tilde{\lambda}}$ will therefore be:

\begin{equation*}
    \mathcal{Q} \left[ \nnu_{\tilde{\lambda}} ; \lambda \right] \ = \  \left( \frac{\partial \tilde{\lambda}}{\partial \lambda} \right)^{2} \mathcal{Q} \left[ \nnu_{\tilde{\lambda}} ; \tilde{\lambda} \right] =  \frac{ \eta}{ ( 1 + \epsilon(1-\eta) \lambda)^{2}}   \frac{1}{ \lambda ( 1 + \eta \lambda + (1-\eta) \epsilon \lambda)} 
\end{equation*}
which is smaller than $\mathcal{Q} \left[ \nnu_{\lambda} ; \lambda \right]$, the QFI of the starting thermal state. Finally, additional information is provided by the very same probability distribution of success for the approximate one-photon subtraction. Specifically, the corresponding FI is:
\begin{equation}
    \mathcal{F} \left[ \{ \wp_{0}, \wp_{1}  \} \right] \ =  \   \frac{ \left( \partial_{\lambda} \wp_{1}  \right)^{2}}{\wp_{1}}  +  \frac{ \left( \partial_{\lambda} \wp_{0} \right)^{2}}{\wp_{0}}
    \ = \ \frac{ \epsilon ( 1 - \eta)}{ \lambda ( 1 + \lambda \epsilon ( 1 - \eta) )^{2}}
\end{equation}
Overall, the full information that we can extract from the approximate one-photon subtraction protocol on $\nnu_{\lambda}$, assuming the best possible detection strategy on the output states, is \cite{weakmeas2015}:
\begin{equation}
\label{eq:TotalInfo}
    \mathcal{F}_{\mathrm{tot}} \ := \  \mathcal{F} [ \{ \wp_{0} , \wp_{1}  \} ] \ + \ \wp_{1} \mathcal{Q} \left[ \rrho_{\lambda} \right] \ + \ \wp_{0} \mathcal{Q} \left[ \rrho^{\oslash}_{\lambda} \right]
\end{equation}
This is plotted against the reference quantity $\mathcal{Q} [ \nnu_{\lambda} ]$ in Fig.(\ref{fig:QFItotal}).

It appears as if all the advantage we found with the realistic one-photon subtracted thermal state is precisely cancelled by the weighted sum in Eq.(\ref{eq:TotalInfo}), so that the total extractable information comes to coincide with the initial QFI of $\nnu_{\lambda}$. This can be understood by noting that $\mathcal{F}_{\mathrm{tot}}$ is necessarily upper-bounded by the QFI of the output state of the beam splitter before the post-selection, Eq.(\ref{eq:RBS1}), since all the collected outcomes (including the one encoded in the acceptance probability) can be viewed as a single string of outcome for a measurement performed on the joined state of the two output modes. But this is simply the result of the BS unitary evolution of the vacuum with the initial state whose temperature we wanted to estimate, therefore the QFI for the temperature parameter will be the same as for the thermal state we started with. \\

One can also derive a lower bound on $\mathcal{F}_{\mathrm{tot}}$, taking advantage of a convexity property of the QFI \cite{QFIconvexity}: if a family $\varrho_{\lambda}$ of quantum states encoding a parameter $\lambda$ is decomposed into a convex mixture according to:
\begin{equation}
    \varrho_{\lambda} \ \ = \ \ \sum_{a} p^{a}_{\lambda} \varrho^{a}_{\lambda}
\end{equation}
where $p^{a}_{\lambda}$ is a probability distribution over the variable $a$, depending upon the parameter $\lambda$, then:
\begin{equation}
\label{eq:QFIconvex}
    \mathcal{Q} \left[ \varrho_{\lambda} \right] \ \ \leq \ \ \sum_{a} p^{a}_{\lambda} \mathcal{Q} \left[ \varrho^{a}_{\lambda} \right] \ + \ \mathcal{F} \left[ \{ p^{a}_{\lambda} \} \right]
\end{equation}
This is called an \emph{extended convexity property} for the QFI, since it states that the Quantum Fisher Information of a mixture is no greater than the average of the QFI of the states in the ensemble \emph{plus} the classical FI of the probability distribution of the ensemble. For us, the convex combination is:
\begin{equation}
   \varrho_{\lambda} \ \ = \ \  \wp_{1}(\lambda) \rrho_{\lambda} \ + \ \wp_{0}(\lambda) \rrho^{\oslash}_{\lambda} \ \ = \ \ \nnu_{\eta \lambda}
\end{equation}
which is simply the statement that if we ignore the outcome of the on-off detector in the realistic one-photon subtraction protocol, the outcome will be again a thermal state, with an average number of photons reduced by the beam-splitter transmittance: $\eta \lambda$. The extended convexity property (\ref{eq:QFIconvex}) then implies:
\begin{equation}
\label{eq:ThConvex}
    \mathcal{Q} \left[ \nnu_{\eta \lambda} ; \lambda \right] \ \ = \ \ \frac{ \eta}{\lambda ( 1 + \eta \lambda)} \ \ \leq \ \ \mathcal{F}_{\mathrm{tot}}
\end{equation}
As $\eta \to 1$, the lower bound converges to the QFI of $\nnu_{\lambda}$, while in general, for $0 < \eta < 1$, we have:
\begin{equation}
    \frac{ \eta}{\lambda ( 1 + \eta \lambda )} \ < \ \frac{1}{\lambda ( 1 + \lambda)} \ = \ \mathcal{Q} \left[ \nnu_{\lambda} \right]
\end{equation}

The fact that the metrological advantage of realistic one-photon subtracted thermal state cannot be actually harnessed with the optimal detection strategy, i.e. photon-number detection, is of minor hindrance from a practical perspective, since photon-number detection is experimentally very challenging anyway. Therefore it is meaningful to explore whether the total extractable information from the realistic one-photon subtracted thermal state can be larger than the FI of the initial state for sub-optimal, but experimentally more relevant detection strategies. \\

\begin{figure}
\centering
\includegraphics[width=0.7 \textwidth]{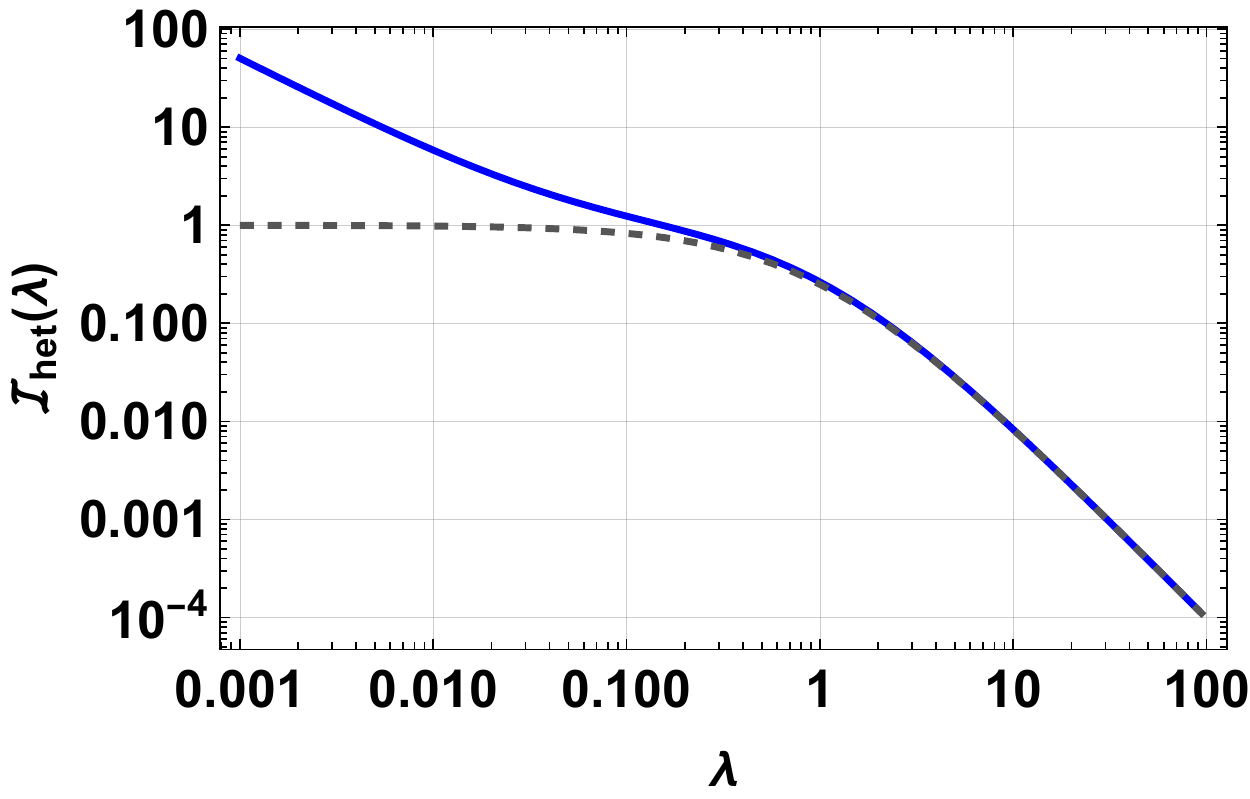}
$\quad $
\caption{\label{fig:HetTot}
Total information retrieved from realistic photon-subtraction protocol followed by heterodyne detection of the output (blue curve) and Fisher Information of the heterodyne distribution of $\nnu_{\lambda}$ for comparison (dashed, gray curve).  The detection efficiency of the on-off measurement is $\epsilon = 0.99$ and the transmittance of the BS is $\eta = 0.95$. }
\end{figure}

\begin{figure}
\centering
\includegraphics[width=0.7 \textwidth]{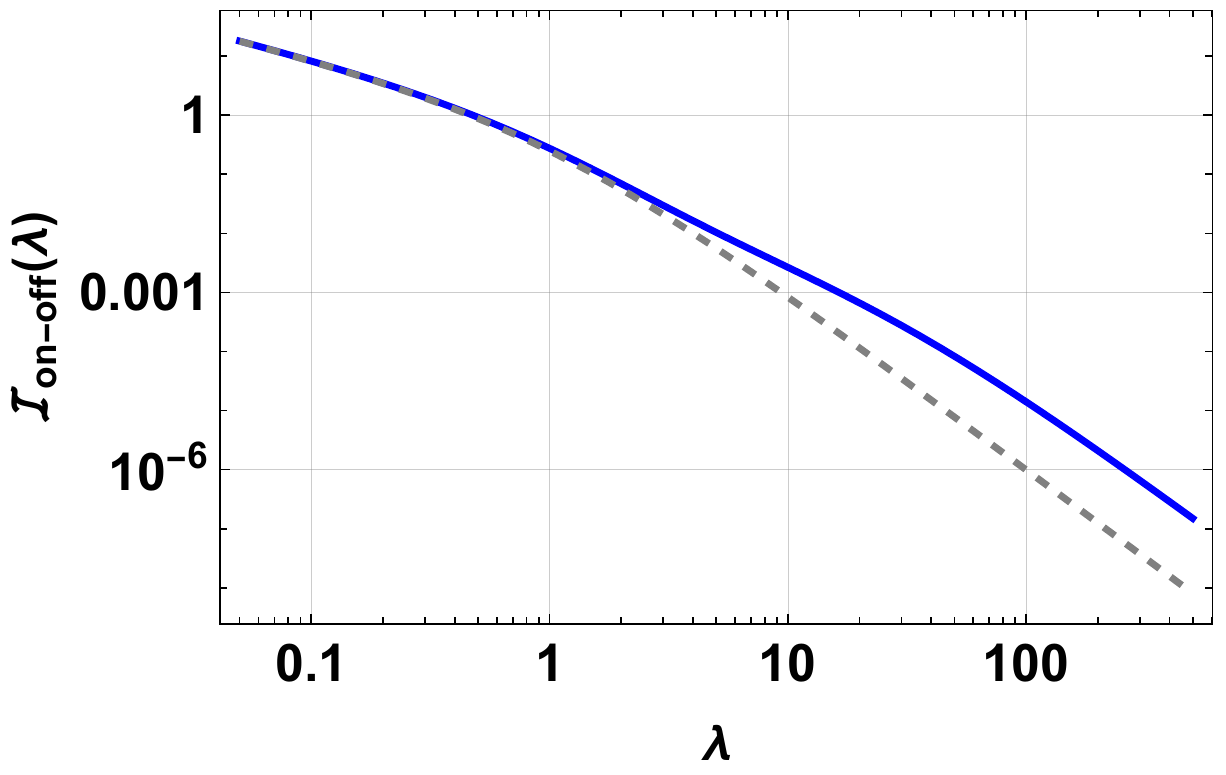}
$\quad $
\caption{\label{fig:YNtot}
Total information retrieved from realistic photon-subtraction protocol followed by on-off detection of the output (blue curve) and Fisher Information of the heterodyne distribution of $\nnu_{\lambda}$ for comparison (dashed, gray curve).  The detection efficiency of the on-off measurements is $\epsilon = 0.99$ and the transmittance of the BS is $\eta = 0.95$. }
\end{figure}

\indent
In Fig.\ref{fig:HetTot} we plotted the total information retrieved by the realistic photon-subtraction scheme, but assuming \emph{heterodyne detection} on the output states of the beam splitter (blue curve), compared to the Fisher Information for the heterodyne distribution of the initial thermal state. For $\lambda \lesssim 0.3$, the post-selection provides a better information rate, even taking into account the photon-subtraction probabilities. Analogously, in Fig.\ref{fig:YNtot}, both for the post-selection scenario (blue curve) and for the one without post-selection (gray, dashed curve), we chose as final measurement an on-off photon detection with the same efficiency $\epsilon = 0.99$. Now the advantage of the conditional strategy is seen for $\lambda \gtrsim 2$.

\section{Information-cost rate}
\begin{figure}
\centering
\includegraphics[width=0.7 \textwidth]{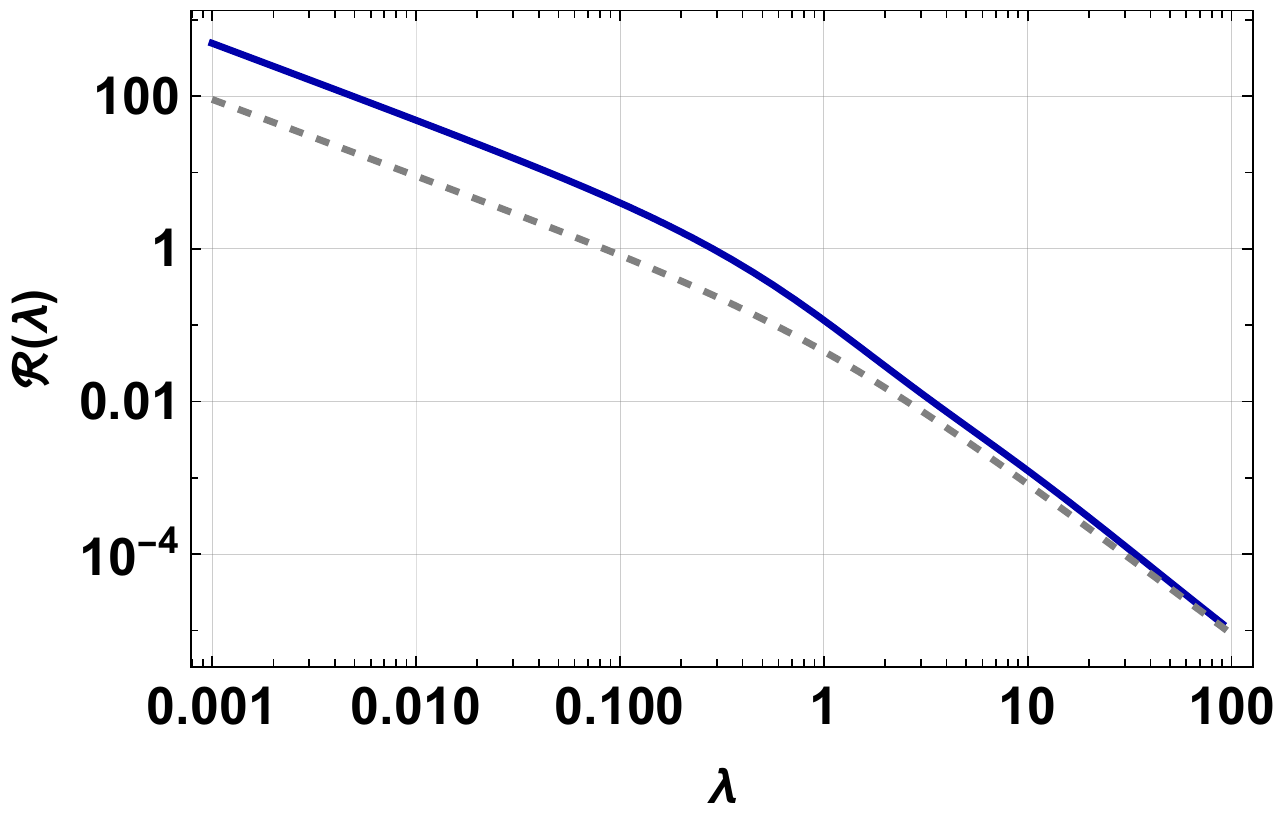}
$\quad $
\caption{\label{fig:Rcomp}
Information-cost ratio for the realistic photon subtraction protocol followed by homodyne detection ($R_{ps}$, solid blue curve) and for the initial thermal state subjected to the optimal photon-number measurement ( $\mathcal{R}_{0}$, dashed gray curve), as functions of the parameter $\lambda$ to be estimated. The cost of preparation was assumed to be $C_{P} = 1$, that of post-selection $C_{S} = 0.5$ and the measurement cost is $C_{M} = 10$. The detection efficiency of the on-off measurements is $\epsilon = 0.99$ and the transmittance of the BS is $\eta = 0.95$. }
\end{figure}

To take into account the different costs of various detection schemes, it is convenient to introduce cost parameters. We will denote by $C_{P}$ the cost of the preparation of the initial quantum state, by $C_{S}$ the cost of the post-selection process (i.e., essentially that of an on-off measurement) and by $C_{M}$ the cost of the final measurement, which is performed only when the post-selection is effective, when realistic photon-subtraction is considered. Whenever the post-selection fails, we will perform the least expensive measurement, which is on-off detection with the same efficiency $\epsilon$, on the resulting state $\rrho^{\oslash}_{\lambda}$. The Fisher Information, with respect to the initial parameter $\lambda$, extractable from the resulting statistics is readily calculated:
\begin{equation}
    \mathcal{F}^{\oslash} \ = \ \frac{ \epsilon \eta}{\lambda ( 1 + \epsilon \lambda )^{2} ( 1 + \epsilon ( 1 - \eta) \lambda ) }  \,.
\end{equation}
The information-cost rate for the post-selection strategy, $\mathcal{R}_{ps}$, is then given by \cite{NYH2020,postselectcost}:
\begin{equation}
    \mathcal{R}_{ps} \ = \ \frac{ \wp_{1} \mathcal{F}_{M} \left[ \rrho_{\lambda} \right] + \wp_{0} \mathcal{F}^{\oslash}  + \mathcal{F} \left[ \{ \wp_{0}, \wp_{1} \}\right] }{ C_{P} + C_{S} + \wp_{1} C_{M} + \wp_{0} C_{S} } \,.
\end{equation}
where $\mathcal{F}_{M} \left[ \rrho_{\lambda} \right]$ is the Fisher Information for the measurement strategy labelled by $M$ performed on the successfully post-selected state $\rrho_{\lambda}$ and we attributed the same post-selection cost $C_{S}$ also to the on-off measurement on the rejected state with the $\wp_{0} C_{S}$ term in the denominator. In the standard quantum estimation protocol, without performing the photon-subtraction, the information-cost rate would be:
\begin{equation}
    \mathcal{R}_{0} \ = \ \frac{ \mathcal{F}_{M} \left[ \nnu_{\lambda} \right]}{ C_{P} + C_{M}  } \,.
\end{equation}
Since $C_{M}$ is usually the largest among the costs and the realistic photon-subtraction strategy pays it only for the more informative states $\rrho_{\lambda}$, it is clear that it is possible to have $\mathcal{R}_{ps} > \mathcal{R}_{0}$. To provide just an example of this fact, let's assume $C_{P} = 1$, $C_{S} = 0.5$ and $C_{M} = 10$. We will take the same $C_{M}$ both for $\mathcal{R}_{ps}$ and $\mathcal{R}_{0}$, but we will assume heterodyne detection on $\rrho_{\lambda}$ and the optimal, photon-number measurement on $\nnu_{\lambda}$ (so that $\mathcal{F}_{M} \left[ \nnu_{\lambda} \right] =  \mathcal{Q} \left[ \nnu_{\lambda} \right]$). This is perhaps unreasonable, since a measurement that has to resolve the full photon-number distribution will be more difficult to perform than heterodyne detection, however it makes our point even stronger, since if $\mathcal{R}_{ps} > \mathcal{R}_{0}$ when the measurement costs are considered the same, it will be true a fortiori when more plausible values are inserted. The comparison is depicted (in logarithmic scale) in Fig.\ref{fig:Rcomp} as a function of $\lambda$, where the blue curve refers to $\mathcal{R}_{0}$ and the dashed gray curve to $\mathcal{R}_{ps}$.

Clearly, at least for low temperature thermal states, the realistic photon-subtraction strategy followed by heterodyne detection has a better information-cost rate than the optimal measurement on the initial thermal state that achieves the Quantum Cramér-Rao bound. The term $\wp_{0} \mathcal{F}^{\oslash}$ in the numerator of $\mathcal{R}_{ps}$ is typically negligible, and one can work with the more compact expression $ \mathcal{R}_{ps} = ( \wp_{1} \mathcal{F}_{M} \left[ \rrho_{\lambda} \right] + \mathcal{F} \left[ \{ \wp_{0}, \wp_{1} \}\right] ) / ( C_{P} + C_{S} + \wp_{1} C_{M} )$. 
On the other hand, it would me meaningless to discard the $\mathcal{F} \left[ \{ \wp_{0}, \wp_{1} \}\right]$ term, since this information is already available anyway. Finally, let us remark that, for $\lambda \ll 1$, there is always a cheap, very effective strategy, which is to perform an on-off measurement directly on $\nnu_{\lambda}$; indeed, such a measurement yields a good approximation to the full photon-number distribution in that limit. However, this strategy is doomed to failure for large enough values of $\lambda$, implying that our post-selection protocol is the most efficient of all the considered ones in a wide range of temperatures.

\section{CONCLUSION AND REMARKS}

In this paper, we explored the effectiveness of single-photon subtraction in enhancing the QFI for the temperature parameter\footnote{To be precise, we referred to the parameter to be estimated as the \emph{temperature} even if we always used the average photon number $\lambda = (e^{\beta} - 1)^{-1}$, while the actual temperature of the thermal state is (proportional to) $\beta^{-1}$. Although the FI for $\lambda$ or $\beta^{-1}$ are related by a temperature-dependent conversion factor dictated by propagation of uncertainty, this is the same factor for all the curves and it cannot influence the comparisons and the relevant conclusions we have drawn.} of a thermal state of a single mode of radiation. After arguing that ideal single-photon subtraction is indeed effective, we considered a practical implementation involving a highly transmitting beam splitter and an on-off detection on the reflected beam; this setup naturally shifts the focus to the field of probabilistic quantum metrology, which deals with weak measurements and post-selection strategies to concentrate the information on the output quantum state before the final measurement. Although no overall enhancement of the QFI persists once all the contributions to the extractable information from the probabilistic protocol are taken into account, this conclusion holds true only if the ability to perform the optimal measurement, required to achieve the QFI, is assumed. Since this request is often too demanding, we examined other measurement protocols on the final states, such as heterodyne and homodyne detection. An actual advantage of these strategies over the straightforward optimal measurement on the initial thermal state is certified by looking at the information-cost ratio, which is given by the total Fisher Information divided by the sum of the costs of all the involved resources (preparation of the initial state, final measurement and, possibly, post-selection). In particular, we showed that, under very conservative values of the costs, the realistic one-photon subtraction followed by heterodyne measurement on the output only when the subtraction was successful, yields an higher information-cost ratio than the standard optimal strategy implied by the Quantum Cramér-Rao bound. \\

\indent
It should be noticed that modern techniques to achieve \emph{deterministic} single-photon subtraction have been devised (see e.g. \cite{Rosenblum2015,Hon16,Du20} and references therein) and, in light of our initial result, it is conceivable that they would allow to go beyond the Quantum Cramér-Rao bound in the temperature estimation of a thermal state of radiation; nonetheless, these procedures are significantly more contrived than the cheap beam-splitter protocol that we assumed, and the costs should be reconsidered accordingly. We conclude by comparing our results to those in \cite{NYH2020,das2021}, where the necessity of a form of nonclassicality to improve the information-cost ratio with post-selection was proven, but assuming a unitary quantum estimation task. Moreover, a careful analysis should be carried out to assess that an actual, deterministic photon-subtraction is attained in the sense that was assumed here.\\

\indent 
 Finally, our analysis complements and contrasts with a recent work \cite{NYH2020} in which the authors prove the necessity of a form of nonclassicality, namely the negativity of a generalized Kirkwood-Dirac distribution, to achieve an improvement of the information-cost ratio with post-selection (but see also \cite{das2021}). In our case, no form of nonclassicality can be at play, because we are always dealing with fully classical states according to Glauber's definition. This does not contradict the results of \cite{NYH2020}, since they considered \emph{unitary} quantum estimation problems and the consequent maximization over the probing state at the input of the unitary channel, while temperature cannot be considered a unitarily-imprinted parameter and no further maximization can be made. Therefore, the present work shows that the restriction in \cite{NYH2020} to unitarily-imprinted parameters is actually necessary to arrive at their conclusion and, with more general assumptions, an improvement of the information-cost ratio with probabilistic metrology is possible also in a fully classically-describable setting. In this respect, we stress that our goal was not to determine the absolute best strategy to estimate the temperature of a thermal light state, but rather to lay down a full example that shows the actual advantages that probabilistic (quantum) metrology can provide even without the need of nonclassical phenomenology.

\newpage 

\bibliographystyle{iopart-num}
\bibliography{GnGphsubt}

\end{document}